\documentclass{article}
\usepackage{amsmath}
\usepackage{amssymb}
\usepackage{amsfonts}
\usepackage[dvips]{graphicx,color}

\begin{document}

\title{Constraints of cluster separability and covariance on current operators}

\author{W. N. Polyzou\\
Department of Physics and Astronomy,\\
The University of Iowa, Iowa City, IA
52242 \\
B. D. Keister\\
NSF Division of Nuclear Physics\\
Arlington, VA 22230
}

\vspace{10mm}

\date{\today}

\maketitle

\begin{abstract}

Realistic models of hadronic systems should be defined by a dynamical
unitary representation of the Poincar\'e group that is also consistent
with cluster properties and a spectral condition.  All three of these
requirements constrain the structure of the interactions.  These
conditions can be satisfied in light-front quantum mechanics,
maintaining the advantage of having a kinematic subgroup of boosts
and translations tangent to a light front.  The most straightforward
construction of dynamical unitary representations of the Poincar\'e
group due to Bakamjian and Thomas fails to satisfy the cluster
condition for more than two particles.  Cluster properties can be
restored, at significant computational expense, using a recursive
method due to Sokolov.  In this work we report on an investigation of
the size of the corrections needed to restore cluster properties in
Bakamjian-Thomas models with a light-front kinematic symmetry.  Our
results suggest that for models based on nucleon and meson degrees of
freedom these corrections are too small to be experimentally observed.

\end{abstract}

\vspace{10mm}

Light-front quantum mechanics \cite{dirac49} has the desirable feature
that it is easy to construct realistic quantum mechanical models that
are exactly Poincar\'e invariant\cite{wigner39}, with a kinematic
subgroup that leaves a light front invariant.  The most
straightforward construction, due to Bakamjian and Thomas
\cite{bakamjian53}\cite{coester82}, achieves this by requiring both
the generators of the kinematic subgroup and the total light-front
spin to be non-dynamical.  A dynamical mass operator is defined by
adding interactions that commute with the kinematic generators and the
kinematic light-front spin to the non-interacting invariant mass.  The
dynamical Poincar\'e generators are well-defined functions of this
mass operator and these kinematical operators.

While the original Bakamjian-Thomas construction was intended for
two-body systems, the construction gives a dynamical representation of
the Poincar\'e group for any number of particles.  However, while this
construction satisfies exact Poincar\'e invariance, the Poincar\'e
generators do not satisfy cluster properties for systems of more than
two particles.

Sokolov \cite{sokolov77} introduced an inductive construction that
starts with Bakamjian-Thomas two-body models and builds a many-body
unitary representation of the Poincar\'e group consistent with cluster
properties.  Sokolov's construction can be formulated to preserve the
light-front \cite{coester82} kinematic subgroup.  The key elements
in the Sokolov construction are unitary transformations that map
tensor products of the Poincar\'e group into S-matrix equivalent
light-front Bakamjian-Thomas representations.  These transformations
preserve the $S$ matrix but they do not preserve cluster properties.

In Sokolov's construction many-body Poincar\'e generators that satisfy
cluster properties are expressed as functions of several of these
unitary transformations and Bakamjian-Thomas interactions.  The
Bakamjian-Thomas generators are recovered in the limit that all of
these unitary transformations become the identity.  For systems of
four or more particles this limit does not preserve the $S$-matrix.

Because of its complexity, there have been no dynamical few-body
calculations that utilize the full Sokolov construction.  On the other
hand there have been many calculations based on the Bakamjian-Thomas
construction, some involving more than two particles.  The size of the
unitary transformations discussed in the previous paragraph relative
to the identity provides one estimate on the size of the terms needed
to restore cluster properties in Bakamjian-Thomas models.

In this work we use a simple model to investigate how close these
unitary transformations are to the identity.  We consider a system
where an electron scatters off of a proton in the presence of a
bound state of a proton and neutron (a deuteron).  We assume that the
deuteron does not interact with the struck proton or electron.  We
treat the three-nucleon system in two ways.  We first represent the
three-body dynamics by a unitary representation of the Poincar\'e
group that is a tensor product of a one-body representation with an
interacting two-body representation.  This representation satisfies
cluster properties by construction.  We also consider a $S$-matrix
equivalent Bakamjian-Thomas representation of the 2+1 system.  This
representation has a non-interacting light-front spin.  The
interactions are constructed to give identical $S$ matrices.  This is
done by taking the same internal two-body interaction and multiplying
by different delta functions that ensure that the interaction in the
three-particle Hilbert space either commutes with the spectator
unitary representation of the Poincar\'e group (tensor product
representation) or the non-interacting spin of the three-body system
(Bakamjian-Thomas representation).  Both models give the same 
$S$-matrix because the internal two-body interactions are identical.  To
study systems with nuclear physics scales we choose the model two-body
mass Casimir operator to be $M_0^2 + 4mV$ \cite{cps}\cite{Keister91}
where $V$ is a Malfliet-Tjon \cite{malfliet69} nucleon-nucleon
interaction, $M_0$ is the non-interacting two-body invariant mass, and
$m$ is the nucleon mass.  This interaction is a sum of an attractive
and short-range repulsive Yukawa interaction that supports a deuteron
bound state.  The existence of Sokolov's unitary operator $A$ relating
these two representations follows from the identity of the two
$S$-matrices as a consequence of a theorem of Ekstein
\cite{ekstein60}.

In order to focus on the essential features associated with
the violation of cluster properties we ignore all particle spins and
we replace the four-vector current that interacts with the electron
current in the one-photon-exchange-approximation with a scalar
current.  The introduction of the electron makes this a four-body
problem.  The relevant feature is that the scalar current is evaluated
between three-nucleon states with different total energy and momenta.

The Bakamjian-Thomas and tensor product
initial and final states are related by
\begin{equation}
A \vert d, p_3, p_{12} \rangle_{bt} =\vert d, p_3, p_{12} \rangle_{tp},
\label{1}
\end{equation}
where $A$ is the Sokolov operator, $bt$ stands for the 
Bakamjain-Thomas states and $tp$ stands for the tensor product states.
For both models we calculate the quantity 
\begin{equation}
F_x(p_3',p_3,p_{12}):=
\int{} _x\langle d, p_3' , p_{12}' \vert J(0) \vert d, p_3, p_{12} \rangle_x
dp_{12}' 
\label{2}
\end{equation}
where $x \in \{bt,tp\}$.  We use the same variables in both models,
even though the variables used in eq.~(\ref{2}) are not the most natural
choice in the Bakamjian-Thomas model. 

In the tensor product model $F_{tp} (\cdots
)$ is independent of $p_{12}$ and only depends on the momentum
transfer $Q=p_3'-p_3$.  This is the expected behavior.  In the
Bakamjian-Thomas model we find a non-trivial dependence on $p_{12}$.
Because of equation (\ref{1}) we can express the correct tensor 
product result in terms of the Bakamjian-Thomas states and the 
unitary operators $A$:
\begin{equation}
F_{tp}(p_3',p_3,p_{12}):=
\int {}_{bt}\langle d, p_3' , p_{12}' \vert A^{\dagger} J(0) 
A \vert d, p_3, p_{12} \rangle_{bt}
dp_{12}' .
\end{equation}
In the limit $A \to I$ this becomes the Bakamjian-Thomas result.  In
figure's 1 and 2 we plot the difference $(F_{tb}-F_{bt})/F_{tp}$ as a
function of $Q= P'-P$ and $p_{12}$, where $P=p_{12}+p_3$.  We consider
frame where the $+$ component of the momentum transfer is zero, which
is always possible for spacelike momentum transfers.  We choose the
light front $z+t=0$, assume that the momentum transfer $Q$ is in the
$x$ direction and investigate the dependence on $p_{12}$ in the $x$
(parallel) or $y$ (perp) directions.

The results are shown in figures 1 and 2.  In both cases the figures
exhibit a non-trivial (and unphysical) dependence on $p_{12}$ in the
Bakamjain-Thomas case.  However, for this problem,  which uses
parameters that have scales expected in nuclear physics models with
meson-exchange interactions, the size of the corrections needed to
restore cluster properties is too small to be measured in laboratory
experiments.  This investigation suggests that it is reasonable 
to construct light-front quantum mechanical models of few-nucleon 
systems using only the Bakamjian-Thomas representation of the Poincar\'e 
group, without including the corrections due to the Sokolov operators.  

Had we instead constructed our deuteron out of sub-nuclear degrees 
of freedom involving stronger binding and larger internal momenta
the size of these corrections could large enough to be observable.

\begin{figure}
\begin{minipage}[t]{7.3cm}
\begin{center}
\includegraphics[width=7.0cm,clip]{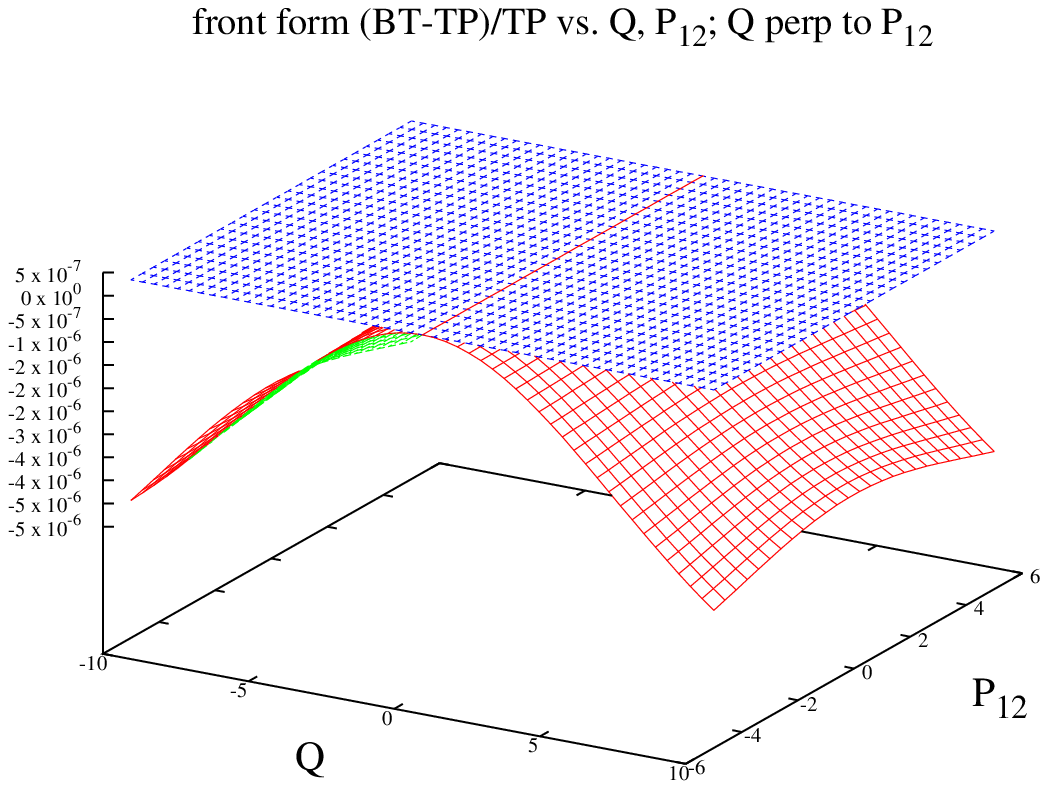}
\caption[Short caption for figure 1]{\label{labelFig1}                          
Front form  - $p_{12}\perp Q$                                                   
}
\end{center}
\label{fig.1}
\end{minipage}
\hfill
\begin{minipage}[t]{7.3cm}
\begin{center}
\includegraphics[width=7.0cm,clip]{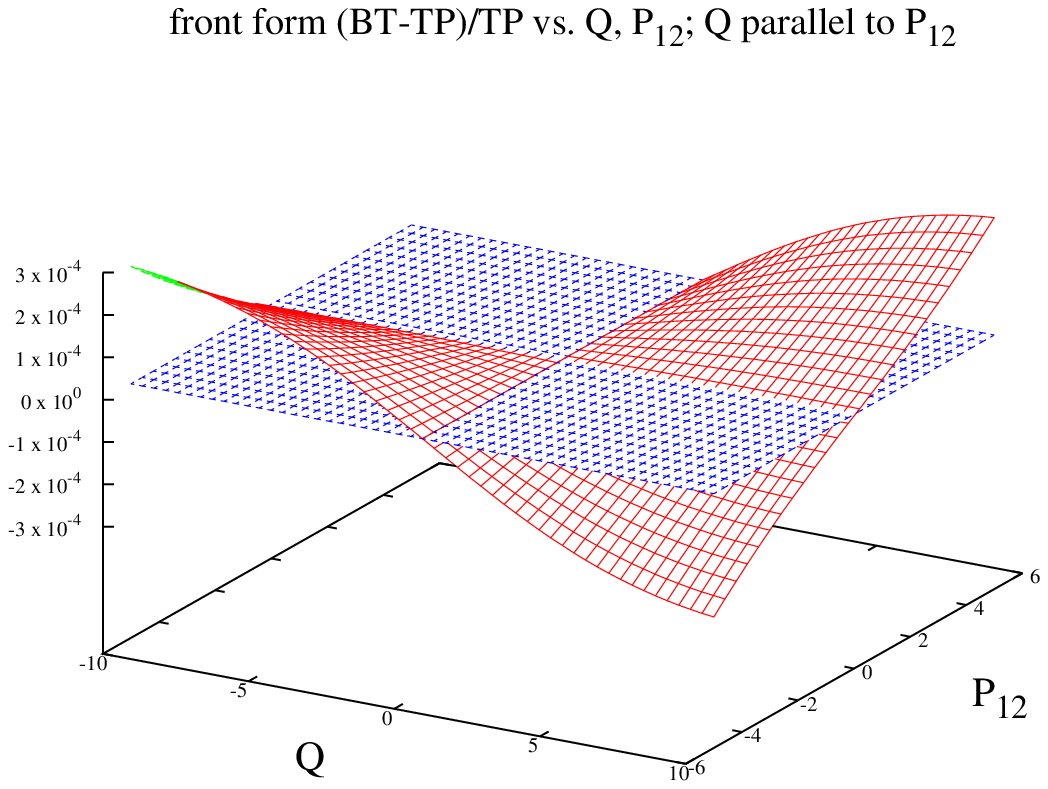}
\caption[Short caption for figure 2]{\label{labelFig2}                          
Front form  - $p_{12} \Vert  Q$                                                 
}
\end{center}
\end{minipage}
\label{fig.2}
\end{figure}

This work supported in part by the U.S. Department of Energy, under 
contract DE-FG02-86ER40286.

\end{document}